\documentstyle[12pt]{article}
\begin{document}
\baselineskip=20pt
\pagestyle{empty}
\begin{flushright}
hep-ph/9706398\\
 MRI-PHY/P970512   
\end{flushright}
\def\rp{R_p}
\def\msd{m^2_{\tilde d}}
\def\msdn{m^2_{{\tilde d}_n}} 
\begin{center}
{\bf {R parity violating contribution to
 $e^+e^-(\mu^+\mu^-)\rightarrow t{\bar c}$}}
\end{center}
\vskip .25in
\begin{center}
 Uma Mahanta and Ambar Ghosal\\
\vskip .1in
 Mehta Research Institute\\
 Chhatnag Road, Jhusi\\    
 Allahabad-221506, India\\
\end{center}
\vskip 1in
\begin{center}
{\bf Abstarct}
\end{center}
\vskip .25in
In this article we consider the contribution 
of $R_p$ violating couplings
to the process  $e^+e^-(\mu^+\mu^-)\rightarrow 
t{\bar c}$ at high energy
lepton collider. We show that the present 
upper bound on the relevant
$R_p$ violating coulpings obtained from 
low energy measurements
 would produce a few hundred to a thousand 
top-charm events at the next linear 
 $e^+e^-(\mu^+\mu^-)$ collider. 
Hence, it should be possible to observe 
the rare process at future lepton collider.\\
\vskip 2in
(To appear in Phys.ReV. D)
\newpage
\pagestyle{plain}
In the Standard Model (SM) flavor changing neutral 
current (FCNC)
processes are suppressed
since they are forbidden at the tree level because 
of the well known
GIM mechanism [1]. The two higgs doublet extension 
of the SM usually
 incorporates a discrete symmetry [2] to guarantee 
tree level natural
 flavor conservation. In the minimal supersymmetric 
standard model
(MSSM) [3] there are two higgs doublets, however 
supersymmetry forces
one of them to couple to $U^c_L$ and the other 
to $D^c_L$ which forbids
 tree level FCNC processes. 
Some FCNC processes  like $b\rightarrow
s\gamma $ [4] get small but measurable 
contributions from heavy fermions in
the loop both in the SM and the MSSM. 
A similar enhancement however is 
not expected in the case of top-charm 
transition. If MSSM is extended by
 the addition of R-parity ($R_p$) violating 
interactions [5-7]  which violate
 either B or L but not both then flavor changing 
neutral current processes
can occur at the tree level. In fact, $\rp$ violating 
couplings because of its
complex flavor structure opens up the 
possibility of many FCNC processes
at the tree level. The effect of $\rp$ 
violating couplings on some of these
FCNC processes have already been 
considered in the literature [8,9,10,11].
The FCNC process involved in top-charm 
production is of special interest
because top quark is the heaviest of 
all fermions and therefore
flavor symmetry violation is expected 
to be maximum for this process.
Recently, the authors of Ref.[10] have 
studied the effects of $\rp$ violating couplings 
on top-charm production at hadron collider. 
The relatively clean environment of a lepton collider 
compared to a hadron collider will make the detection 
of the signal for the rare process easier.
In this article we shall consider the 
contribution of the L and $\rp$
violating interaction 
$L_I=-\lambda^{\prime}_{ijk}({\tilde d}^k_R)^* 
 ({\bar e}^i_L)^c u^j_L + h.c. $ to the
 proces 
$e^+e^-(\mu^+\mu^-)\rightarrow t{\bar c}$
via the exchange of down squarks 
in the u channel. In the SM 
 the process  
$e^+e^-(\mu^+\mu^-)\rightarrow t{\bar c}$
proceeds at one loop 
level via  $e^+e^-(\mu^+\mu^-)\rightarrow \gamma^*,
Z^*\rightarrow t{\bar c}$. 
The effective (off shell)
$\gamma t{\bar c}$ and
$Zt{\bar c}$ vertices can be 
evaluated to one loop and used in the
 calculation. The production rate for 
this rare process has been
calculated in the framework of SM and is 
predicted to be extremely
small [12]. One reason being 
in SM $t{\bar c}$ production, 
unlike $b{\bar s}$
production, does not get a large 
contribution from heavy 
fermion in the loop.
The MSSM contribution to the 
branching ratio for 
$e^+e^-\rightarrow t{\bar c}$
has also been calculated [13]
and has been shown to be small 
compared to that
from the SM. Hence any experimental 
detection of 
this process beyond 
the SM prediction will therefore point to the 
existence of new physics
 other than MSSM.\hfill
\vskip .1in
\noindent
 The above $R_p$ violating interaction 
  has been written in the quark mass basis. 
It should
however be noted that the squark and
 quark mass matrices are not diagonal 
in the same basis [14].
 Let ${\tilde D}$
be the 6$\times $6 mass matrix 
that diagonalizes 
the down squark mass matrix
$M^2_{\tilde d}$ and 
$G_{ik}$ be the 6$\times $3 
matrix that relates the
weak gauge edigenstates
${\tilde d}^k_R$ to the mass 
eigenstates ${\tilde d}^m$ 
(${\tilde d}^k_R={\tilde d}^m G_{mk}$, 
$m$=1-6 and $k$=1-3).
The transition 
matrix for the 
process $e^+e^-\rightarrow t{\bar c}$ 
at the tree level
 is given by
\begin{equation}
M=\sum_{k,l,n} { (\lambda^{\prime}_{12k} \lambda^{*\prime}_{13l})
\over 2 (u-\msdn )}G_{nl}G^*_{nk}
{\bar u}_{tL}(p_1^{\prime})\gamma_{\mu} v_{cL}(p_2^{\prime})
{\bar v}_{eL}(p_2)
\gamma^{\mu} u_{eL}(p_1). 
\end{equation}
\noindent
$u_{eL}(p_1)$ and ${\bar v}_{eL}(p_2)$ are the incoming 
spinors for $e^-$
and $e^+$. ${\bar u}_{tL}(p_1^{\prime})$ 
and $v_{cL}(p_2^{\prime})$ are 
the outgoing spinors for t and ${\bar c}$. 
$m_{{\tilde d}_n}$ 
is a generic
 mass of the
 down squarks and $u=(p_1-p_2^{\prime})^2=
(p_1^{\prime}-p_2)^2$.\hfill
\vskip .1in
\noindent
If we make the simlifying assumption that 
the down squark mass eigenstates
are degenerate then the expression for M reduces to
\begin{equation}
M=\sum_k {(\lambda^{\prime}_{12k} \lambda^{*\prime}_{13k})
\over 2 (u-\msd )}
{\bar u}_{tL}(p_1^{\prime})\gamma_{\mu} v_{cL}(p_2^{\prime})
{\bar v}_{eL}(p_2)
\gamma^{\mu} u_{eL}(p_1).
\end{equation}
\noindent
This assumption is however is not very 
realistic since one of the down 
squark mass eigenstates usually tends to 
be much lighter than the others
 because of the large radiative corrections 
to the down squark mass
matrix induced by the top quark. However  the upper bounds
on  $R_p$ violating couplings has been derived in 
Ref. 8 assuming
 that the down squarks are degenerate and we shall 
therefore stick to it
in the present work. The determination of the bounds 
on $R_p$ violating
couplings from low energy experiments,
in the case where the down squarks are not degenerate and there
is considerable mixing between them, is rather complicated.

 We shall assume that the operating CM energy ${\sqrt s}$ 
of the next linear
 $e^+e^-$ collider is large enough
 (${\sqrt s}$=500 GeV or 1000 GeV) so that we can 
ignore the  mass of incoming
$e^+,e^-$ and outgoing ${\bar c}$. The square of the 
invariant transition
matrix element for Left-handed (LH) $e^-$ 
incident on 
unpolarized $e^+$ is given by
\begin{equation}
 |M|^2=2N_c{( \lambda^{\prime}_{12k} \lambda^{*\prime}_{13k})
( \lambda^{*\prime}_{12l} \lambda^{\prime}_{13l})
\over  (u-\msd )^2}(p_1.p_2^{\prime})(p_1^{\prime}.p_2)
\end{equation}
\vskip .1in
\noindent
In the above the repeated indices $k$ and $l$ 
are assumed to be summed over.
For ${\sqrt s}\gg m_c,m_e$ we can make use of 
the approximate relation
$(p_1.p_2^{\prime})(p_1^{\prime}.p_2)\approx {u(u-m^2_t)\over 4}$. 
Note that
the numerator does not contain any term linear in the 
incoming or outgoing
 fermion masses. Such terms can only arise from the 
interfernce between
two currents of opposite chirality which is 
absent in our case.
 The differential scattering
 cross-section for LH $e^-$ incident on 
unpolarized $e^+$ turns out to be
\begin{equation}
{d\sigma\over dt}= {N_c\over 32 \pi s^2} 
\vert\lambda^{\prime}_{12k}
 \lambda^{*\prime}_{13k}\vert ^2{u(u-m^2_t)\over (u-\msd)^2}
\end{equation}
\noindent
where $t=-{1\over 2}(s-m^2_t)(1-\cos\theta)$ and 
 $u=-{1\over 2}(s-m^2_t)(1+\cos\theta)$. 
$\theta$ is the angle between
 the incoming $e^-$ and outgoing t quark in the CM frame.
The repeated index k is assumed to be summed over.
Integrating over all angles the total 
cross-section for $e^+e^-\rightarrow
t{\bar c}$ is given by
\begin{equation}
\sigma  (s)={N_c\over 32 \pi s^2} 
\vert\lambda^{\prime}_{12k}
 \lambda^{*\prime}_{13k}\vert ^2[(s-2m^2_t)$$
$$+{\msd s\over s+\msd-m^2_t}
-(2\msd-m^2_t)\ln{s+\msd-m^2_t\over \msd}].
\end{equation}
\noindent
The most stringent bound on $ \vert\lambda^{\prime}_{12k}
 \lambda^{*\prime}_{13k}\vert ^2$ follows 
from low energy measurements of 
$K^+\rightarrow \pi^+\nu{\bar\nu}$ [15], 
$b\rightarrow s\nu{\bar \nu}$ [15]
and $\nu_e$ mass [16]. The bound obtained in 
this way depends on the value of
$m_{\tilde d}$. For $m_{\tilde d}\approx $ 200 GeV 
the upper bound [8] is
given by  
$\vert\lambda^{\prime}_{12k}\lambda^{*\prime}_{13k}\vert ^2\le
3.4\times 10^{-4}$. 
The bound  ($\vert\lambda^{\prime}_{12k}
\lambda^{*\prime}_{13k}\vert ^2\le .0512$)
that follows from flavor conserving processes
like $A^e_{FB}$ and atomic parity violation is 
much weaker [5,17]. For $m_{\tilde d}
\approx 200 $ GeV, $m_t\approx$ 175 GeV 
and $N_c\approx$ 3 the cross-section 
$\sigma (e^+e^-\rightarrow t{\bar c})$
turns out to be 8.77 (3.32) fb 
at ${\sqrt s}=500 (1000)$ GeV. With an
 integrated luminosity of 50 (300) fb$^{-1}$ 
[18] 
per year at 
 ${\sqrt s}=500 (1000)$ GeV and a detection 
efficiency of 70\% 
about 310 (700) $t{\bar c}$ events are 
expected at an $e^+e^-$ collider.
The effect of QCD corrections to leading 
order will increase the cross-section
 by a factor of $(1+c {\alpha_3(s)\over \pi})$ 
where c is a number of order
unity. However since ${\alpha_3(s)\over \pi}$ is 
rather small for
${\sqrt s} \approx 500 (1000)$ GeV the effect of 
QCD corrections will
be almost inappreciable. At LEP2 where 
unpolarized $e^+$ and $e^-$ beams will
collide at ${\sqrt s}=200 $ GeV the cross-section for 
$e^+e^-\rightarrow t{\bar c} $ is .9 fb 
which is only one tenth of its value
at ${\sqrt s}=500$ GeV.
 In fig. 1 we have plotted the cross-section for
the process
$e^+e^-\rightarrow t{\bar c}$ as a function of $m_{\tilde d}$ for
${\sqrt s}=500 $ GeV and 1000 GeV, keeping 
 $\vert\lambda^{\prime}_{12k}\lambda^{*\prime}_{13k}\vert ^2$ 
fixed at
$3.4\times 10^{-4}$. We find that 
even for $m_{\tilde d}\approx 600$ GeV 
the cross-section is large enough 
to produce about 35 $t{\bar c}$ events
at ${\sqrt s}=500$ GeV.\hfill
\vskip .1in
\noindent
A high energy $\mu^+\mu^-$ collider would 
also provide a relatively
clean environment to look for flavor 
changing top-charm events. For 
 $m_{\tilde d}\approx 200$ GeV the upper bound on 
 $\vert\lambda^{\prime}_{22k}\lambda^{*\prime}_{23k}\vert ^2$ 
is given by 
$7.5\times 10^{-4}$. Hence a $\mu^+\mu^-$ linear 
collider operating at
${\sqrt s} =500 (1000) $ GeV with an integrated 
luminosity of 50 (300) 
fb$^{-1}$ [19] and a 70\% detection 
efficiency is expected to produce
about 665 (1480) $t{\bar c}$ events. 
Note that  although the process
$e^+e^-(\mu^+\mu^-)\rightarrow t{\bar c}$ 
has been analysed here from the 
standpoint of $R_p$ violating interactions 
the same could also arise from
flavor violating leptoquark interactions. 
Top charm production at a high energy
$e^+e^-$ or $\mu^+\mu^-$ collider has 
several phenomenolgical
vantage points. First the 
experimental signature of the final state
 is extremely clean, a fat jet 
recoiling against a relatively thin jet. 
Second the relatively clean environment 
of a lepton collider causes the 
background to be small and enhances the 
detection efficiency of the signal.
The upshot of this discussion is that the 
present low enegy bounds on
$R_p$ violating couplings would produce a 
sufficient number of top charm
events at a high energy lepton collider
that could be easily detected.

A null result on top-charm production would 
be somewhat discouraging since we would be 
losoing good oppertunity to learn about 
flavour changing neutral current events in the 
up-type quark sector. However, it could be used to 
derive an upper bound on  
$\vert\lambda^{\prime}_{12k}
\lambda^{*\prime}_{13k}\vert ^2$. A precise estimation 
of the bound depends crucially on the possible 
background events. Consider the decay of the $t{\bar c}$ 
system: $t + {\bar c}\rightarrow W + b + {\bar c}$.
Now the W boson can decay 
in the leptonic mode through $W\rightarrow l + \nu$ or 
in the hadronic channel as $W\rightarrow q + 
{\bar {q^\prime}}$. For the charged lepton mode, 
the signal contains a high energy charged lepton, 
a large missing energy from the $\nu$, a b-jet and a c-jet.
 The b-jet can be tagged in the final state but the c-jet 
can be mimicked in the light quarks. On the other hand, 
for the hadronic channel, the signal contains 4-jet events. 
The same signal can also be produced through 
$e^+e^-\rightarrow W^+ W^-$ followed by the decay of one $W$ 
in the leptonic mode  and the other in the hadronic mode. 
Alternatively, both the W's can decay in the hadronic mode.
The $W^+ W^-$ production cross-section [20]  is 
7.38(2.83) pb at $\sqrt s$ = 500(1000) GeV and the branching 
ratio of $W^-\rightarrow b{\bar c}$ is $\sim$ .0007. Hence,
the background events arising out of $W^+ W^-$ 
production will be highly suppressed compared to the 
signal.     
On the contrary, the other possible source of background 
arising from 
$e^+e^-\rightarrow b{\bar b}$ followed by the decay 
$b\rightarrow c l \nu$ has a quite large cross-section 
$\sim$ 0.1 pb (.01 pb) at 
$\sqrt s$ = 500(1000) GeV. 
With a luminosity of 50(300) $fb^{-1}$ at 
$\sqrt s$ = 500(1000) GeV the signal to background 
ratio is ${S\over{\sqrt{S +B}}}\approx$ 1.6(4.5). 
The energy of the final state charged lepton 
produced from the decay of b-quarks is much less 
($m_b - m_c$) than the lepton produced from the decay of 
W boson ($\sim m_W$). By applying suitable energy 
cuts on the final state charged lepton it is possible to 
isolate the signal from the background.  
\vskip .1in
\par
In conclusion in this work we have shown 
that the present low energy bound on
$R_p$ violating couplings would produce a 
few hundred to a thousand top-charm
events at the next linear $e^+e^-(\mu^+\mu^-)$ 
collider. 
Given the clear environment of lepton 
collider it should be possible to observe 
this rare FCNC process. 
\newpage
\begin{center}
{\bf References}
\end{center}
\begin{enumerate}

\item S. L. Glashow, J. Iliopoulos and L. Maiani, 
Phys. Rev. D 2, 1285
(1970).

\item S. L. Glashow and S. Weinberg, Phys. Rev. D 15, 
1958 (1977).

\item For reviews of supersymmetry  and 
supersymmetry phenomenology,
see H. P. Nilles, Phys. Rep. 110, 1 (1984); H.E. Haber 
and G. L. Kane, ibid
117, 75 (1985). 

\item R. Ammar et al., Phys. Rev. Lett. 71, 674 
(1993); M. S. Alam
et al., Phys. Rev. Lett. 74, 2885 (1995).

\item V. Barger, G. F. Giudice and T. Han, Phys. 
Rev. D 40, 2987 (1989).

\item S. Dawson, Nucl. Phys. B, 261, 297 (1985).

\item R. Mohapatra, Phys. Rev. D 34, 3475 (1986).

\item K. Agashe and M. Graesser, Phys. Rev. 
D54, 4445, (1996).

\item L.J. Hall and M.Suzuki, Nucl. Phys. B231, 419 (1984),
R.Barbieri and A. Masiero , Nucl. Phys. B267, 679 (1986), 
J. Ellis, et al., Phys.Lett. B150, 142 (1985), G.G.Ross 
and J.W.F.Valle, Phys. Lett. B151, 375 (1985).

\item J.M.Yang, B.L.Young and X.Zhang, hep-ph/9705341.

\item G.Bhattacharya, Nucl. Phys. B (Proc.Suppl.) 52A, 
83 (1997), H. Dreiner, hep-ph/9707442.

\item A. Axelrod, Nucl. Phys. B 209,349 (1982); 
M. Clements et al.,
 Phys. Rev. D 27, 579 (1983); 
V. Ganapathi et al., Phys. Rev. D 27, 579, 
(1983).

\item M. J. Duncan, Phys. Rev. D 31, 1139 (1985); 
B. Mukhopadhyaya and 
A. Raychaudhuri, Phys. Rev. D 39, 280 (1989).

\item J. F. Donoghue, H. P. Nilles and D. Wyler, 
Phys. Lett. B, 128, 55 
(1983).

\item Y. Grossman, Z.Ligeti and E.Nardi, Nucl.Phys. B465, 369 (1996).

\item R. M.Godbole, P.Roy and X.Tata, Nucl. Phys. B401, 67 (1993).

\item S. Davidson, D.Bailey and B.A.Campbell, 
Z. Phys. C61 , 613 (1994), hep-ph/9309310, C.S.Wood et al.,
Science, 275, 1759 (1997), W.J. Marciano, J.L. 
Rosener, Phys. 
Rev. Lett. 65, 2963, (1990), Erratum- $\it{ibid}$. 68, 898 (1992).
 
\item Proceedings of the Conference on Physics 
and Experiments with
 Linear Colliders, Waikoloa, Hawai, USA, 1993.

\item V. Barger, M. S. Berger, J. F. Gunion 
and T. Han, MAD PHY-96-939.

\item W. Beenakker and A. Denner, Int. Jour. Mod. Phys. A9, 4837, (1994).
\end{enumerate}
\newpage
\begin{center}
{\bf Figure Captions}
\end{center}

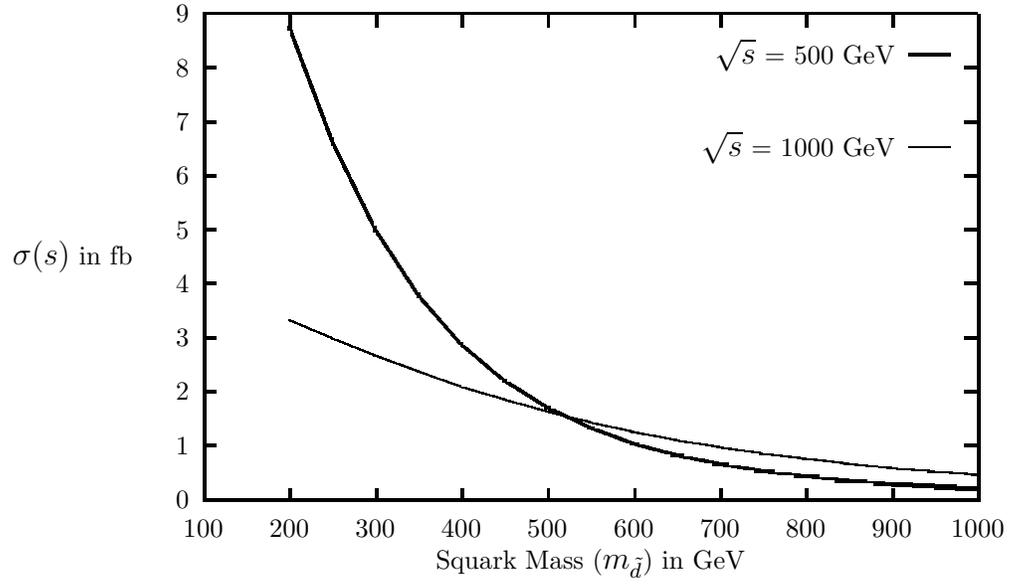
 Fig 1. Cross-section for $e^+e^-\rightarrow t{\bar c}$ 
plotted
against $m_{\tilde d}$ at ${\sqrt s}$ =500 (1000) GeV.
\newpage
\setlength{\unitlength}{0.240900pt}
\ifx\plotpoint\undefined\newsavebox{\plotpoint}\fi
\sbox{\plotpoint}{\rule[-0.500pt]{1.000pt}{1.000pt}}%
\begin{picture}(1500,900)(0,0)
\font\gnuplot=cmr10 at 10pt
\gnuplot
\sbox{\plotpoint}{\rule[-0.500pt]{1.000pt}{1.000pt}}%
\put(220.0,113.0){\rule[-0.500pt]{292.934pt}{1.000pt}}
\put(220.0,113.0){\rule[-0.500pt]{4.818pt}{1.000pt}}
\put(198,113){\makebox(0,0)[r]{0}}
\put(1416.0,113.0){\rule[-0.500pt]{4.818pt}{1.000pt}}
\put(220.0,198.0){\rule[-0.500pt]{4.818pt}{1.000pt}}
\put(198,198){\makebox(0,0)[r]{1}}
\put(1416.0,198.0){\rule[-0.500pt]{4.818pt}{1.000pt}}
\put(220.0,283.0){\rule[-0.500pt]{4.818pt}{1.000pt}}
\put(198,283){\makebox(0,0)[r]{2}}
\put(1416.0,283.0){\rule[-0.500pt]{4.818pt}{1.000pt}}
\put(220.0,368.0){\rule[-0.500pt]{4.818pt}{1.000pt}}
\put(198,368){\makebox(0,0)[r]{3}}
\put(1416.0,368.0){\rule[-0.500pt]{4.818pt}{1.000pt}}
\put(220.0,453.0){\rule[-0.500pt]{4.818pt}{1.000pt}}
\put(198,453){\makebox(0,0)[r]{4}}
\put(1416.0,453.0){\rule[-0.500pt]{4.818pt}{1.000pt}}
\put(220.0,537.0){\rule[-0.500pt]{4.818pt}{1.000pt}}
\put(198,537){\makebox(0,0)[r]{5}}
\put(1416.0,537.0){\rule[-0.500pt]{4.818pt}{1.000pt}}
\put(220.0,622.0){\rule[-0.500pt]{4.818pt}{1.000pt}}
\put(198,622){\makebox(0,0)[r]{6}}
\put(1416.0,622.0){\rule[-0.500pt]{4.818pt}{1.000pt}}
\put(220.0,707.0){\rule[-0.500pt]{4.818pt}{1.000pt}}
\put(198,707){\makebox(0,0)[r]{7}}
\put(1416.0,707.0){\rule[-0.500pt]{4.818pt}{1.000pt}}
\put(220.0,792.0){\rule[-0.500pt]{4.818pt}{1.000pt}}
\put(198,792){\makebox(0,0)[r]{8}}
\put(1416.0,792.0){\rule[-0.500pt]{4.818pt}{1.000pt}}
\put(220.0,877.0){\rule[-0.500pt]{4.818pt}{1.000pt}}
\put(198,877){\makebox(0,0)[r]{9}}
\put(1416.0,877.0){\rule[-0.500pt]{4.818pt}{1.000pt}}
\put(220.0,113.0){\rule[-0.500pt]{1.000pt}{4.818pt}}
\put(220,68){\makebox(0,0){100}}
\put(220.0,857.0){\rule[-0.500pt]{1.000pt}{4.818pt}}
\put(355.0,113.0){\rule[-0.500pt]{1.000pt}{4.818pt}}
\put(355,68){\makebox(0,0){200}}
\put(355.0,857.0){\rule[-0.500pt]{1.000pt}{4.818pt}}
\put(490.0,113.0){\rule[-0.500pt]{1.000pt}{4.818pt}}
\put(490,68){\makebox(0,0){300}}
\put(490.0,857.0){\rule[-0.500pt]{1.000pt}{4.818pt}}
\put(625.0,113.0){\rule[-0.500pt]{1.000pt}{4.818pt}}
\put(625,68){\makebox(0,0){400}}
\put(625.0,857.0){\rule[-0.500pt]{1.000pt}{4.818pt}}
\put(760.0,113.0){\rule[-0.500pt]{1.000pt}{4.818pt}}
\put(760,68){\makebox(0,0){500}}
\put(760.0,857.0){\rule[-0.500pt]{1.000pt}{4.818pt}}
\put(896.0,113.0){\rule[-0.500pt]{1.000pt}{4.818pt}}
\put(896,68){\makebox(0,0){600}}
\put(896.0,857.0){\rule[-0.500pt]{1.000pt}{4.818pt}}
\put(1031.0,113.0){\rule[-0.500pt]{1.000pt}{4.818pt}}
\put(1031,68){\makebox(0,0){700}}
\put(1031.0,857.0){\rule[-0.500pt]{1.000pt}{4.818pt}}
\put(1166.0,113.0){\rule[-0.500pt]{1.000pt}{4.818pt}}
\put(1166,68){\makebox(0,0){800}}
\put(1166.0,857.0){\rule[-0.500pt]{1.000pt}{4.818pt}}
\put(1301.0,113.0){\rule[-0.500pt]{1.000pt}{4.818pt}}
\put(1301,68){\makebox(0,0){900}}
\put(1301.0,857.0){\rule[-0.500pt]{1.000pt}{4.818pt}}
\put(1436.0,113.0){\rule[-0.500pt]{1.000pt}{4.818pt}}
\put(1436,68){\makebox(0,0){1000}}
\put(1436.0,857.0){\rule[-0.500pt]{1.000pt}{4.818pt}}
\put(220.0,113.0){\rule[-0.500pt]{292.934pt}{1.000pt}}
\put(1436.0,113.0){\rule[-0.500pt]{1.000pt}{184.048pt}}
\put(220.0,877.0){\rule[-0.500pt]{292.934pt}{1.000pt}}
\put(15,495){\makebox(0,0){$\sigma(s)$ in fb}}
\put(828,13){\makebox(0,0){Squark Mass ($m_{\tilde d}$) in GeV}}
\put(220.0,113.0){\rule[-0.500pt]{1.000pt}{184.048pt}}
\put(1306,812){\makebox(0,0)[r]{$\sqrt{s}$ = 500 GeV}}
\put(1328.0,812.0){\rule[-0.500pt]{15.899pt}{1.000pt}}
\put(355,857){\usebox{\plotpoint}}
\multiput(356.83,844.85)(0.499,-1.337){128}{\rule{0.120pt}{2.926pt}}
\multiput(352.92,850.93)(68.000,-175.926){2}{\rule{1.000pt}{1.463pt}}
\multiput(424.83,665.47)(0.499,-1.020){126}{\rule{0.120pt}{2.295pt}}
\multiput(420.92,670.24)(67.000,-132.237){2}{\rule{1.000pt}{1.147pt}}
\multiput(491.83,530.61)(0.499,-0.761){128}{\rule{0.120pt}{1.779pt}}
\multiput(487.92,534.31)(68.000,-100.307){2}{\rule{1.000pt}{0.890pt}}
\multiput(559.83,428.19)(0.499,-0.570){126}{\rule{0.120pt}{1.399pt}}
\multiput(555.92,431.10)(67.000,-74.096){2}{\rule{1.000pt}{0.700pt}}
\multiput(625.00,354.68)(0.591,-0.499){106}{\rule{1.443pt}{0.120pt}}
\multiput(625.00,354.92)(65.005,-57.000){2}{\rule{0.721pt}{1.000pt}}
\multiput(693.00,297.68)(0.791,-0.498){76}{\rule{1.845pt}{0.120pt}}
\multiput(693.00,297.92)(63.170,-42.000){2}{\rule{0.923pt}{1.000pt}}
\multiput(760.00,255.68)(1.057,-0.497){56}{\rule{2.375pt}{0.120pt}}
\multiput(760.00,255.92)(63.071,-32.000){2}{\rule{1.188pt}{1.000pt}}
\multiput(828.00,223.68)(1.358,-0.496){42}{\rule{2.970pt}{0.120pt}}
\multiput(828.00,223.92)(61.836,-25.000){2}{\rule{1.485pt}{1.000pt}}
\multiput(896.00,198.68)(1.871,-0.495){28}{\rule{3.972pt}{0.119pt}}
\multiput(896.00,198.92)(58.755,-18.000){2}{\rule{1.986pt}{1.000pt}}
\multiput(963.00,180.68)(2.463,-0.492){20}{\rule{5.107pt}{0.119pt}}
\multiput(963.00,180.92)(57.400,-14.000){2}{\rule{2.554pt}{1.000pt}}
\multiput(1031.00,166.68)(3.129,-0.489){14}{\rule{6.341pt}{0.118pt}}
\multiput(1031.00,166.92)(53.839,-11.000){2}{\rule{3.170pt}{1.000pt}}
\multiput(1098.00,155.68)(4.525,-0.481){8}{\rule{8.750pt}{0.116pt}}
\multiput(1098.00,155.92)(49.839,-8.000){2}{\rule{4.375pt}{1.000pt}}
\multiput(1166.00,147.69)(5.246,-0.475){6}{\rule{9.821pt}{0.114pt}}
\multiput(1166.00,147.92)(46.615,-7.000){2}{\rule{4.911pt}{1.000pt}}
\multiput(1233.00,140.71)(10.507,-0.424){2}{\rule{13.850pt}{0.102pt}}
\multiput(1233.00,140.92)(39.254,-5.000){2}{\rule{6.925pt}{1.000pt}}
\put(1301,133.92){\rule{16.140pt}{1.000pt}}
\multiput(1301.00,135.92)(33.500,-4.000){2}{\rule{8.070pt}{1.000pt}}
\put(1368,129.92){\rule{16.381pt}{1.000pt}}
\multiput(1368.00,131.92)(34.000,-4.000){2}{\rule{8.191pt}{1.000pt}}
\sbox{\plotpoint}{\rule[-0.175pt]{0.350pt}{0.350pt}}%
\put(1306,667){\makebox(0,0)[r]{$\sqrt{s}$ = 1000 GeV}}
\put(1328.0,667.0){\rule[-0.175pt]{15.899pt}{0.350pt}}
\put(355,395){\usebox{\plotpoint}}
\multiput(355.00,394.02)(1.189,-0.501){55}{\rule{0.908pt}{0.121pt}}
\multiput(355.00,394.27)(66.115,-29.000){2}{\rule{0.454pt}{0.350pt}}
\multiput(423.00,365.02)(1.261,-0.501){51}{\rule{0.956pt}{0.121pt}}
\multiput(423.00,365.27)(65.016,-27.000){2}{\rule{0.478pt}{0.350pt}}
\multiput(490.00,338.02)(1.384,-0.501){47}{\rule{1.039pt}{0.121pt}}
\multiput(490.00,338.27)(65.842,-25.000){2}{\rule{0.520pt}{0.350pt}}
\multiput(558.00,313.02)(1.422,-0.501){45}{\rule{1.065pt}{0.121pt}}
\multiput(558.00,313.27)(64.790,-24.000){2}{\rule{0.532pt}{0.350pt}}
\multiput(625.00,289.02)(1.742,-0.501){37}{\rule{1.277pt}{0.121pt}}
\multiput(625.00,289.27)(65.348,-20.000){2}{\rule{0.639pt}{0.350pt}}
\multiput(693.00,269.02)(1.809,-0.501){35}{\rule{1.322pt}{0.121pt}}
\multiput(693.00,269.27)(64.257,-19.000){2}{\rule{0.661pt}{0.350pt}}
\multiput(760.00,250.02)(2.061,-0.502){31}{\rule{1.487pt}{0.121pt}}
\multiput(760.00,250.27)(64.913,-17.000){2}{\rule{0.744pt}{0.350pt}}
\multiput(828.00,233.02)(2.348,-0.502){27}{\rule{1.674pt}{0.121pt}}
\multiput(828.00,233.27)(64.525,-15.000){2}{\rule{0.837pt}{0.350pt}}
\multiput(896.00,218.02)(2.689,-0.502){23}{\rule{1.891pt}{0.121pt}}
\multiput(896.00,218.27)(63.074,-13.000){2}{\rule{0.946pt}{0.350pt}}
\multiput(963.00,205.02)(3.259,-0.502){19}{\rule{2.251pt}{0.121pt}}
\multiput(963.00,205.27)(63.328,-11.000){2}{\rule{1.126pt}{0.350pt}}
\multiput(1031.00,194.02)(3.556,-0.503){17}{\rule{2.433pt}{0.121pt}}
\multiput(1031.00,194.27)(61.951,-10.000){2}{\rule{1.216pt}{0.350pt}}
\multiput(1098.00,184.02)(4.607,-0.504){13}{\rule{3.062pt}{0.121pt}}
\multiput(1098.00,184.27)(61.644,-8.000){2}{\rule{1.531pt}{0.350pt}}
\multiput(1166.00,176.02)(4.539,-0.504){13}{\rule{3.019pt}{0.121pt}}
\multiput(1166.00,176.27)(60.734,-8.000){2}{\rule{1.509pt}{0.350pt}}
\multiput(1233.00,168.02)(5.352,-0.504){11}{\rule{3.487pt}{0.121pt}}
\multiput(1233.00,168.27)(60.762,-7.000){2}{\rule{1.744pt}{0.350pt}}
\multiput(1301.00,161.02)(7.861,-0.507){7}{\rule{4.778pt}{0.122pt}}
\multiput(1301.00,161.27)(57.084,-5.000){2}{\rule{2.389pt}{0.350pt}}
\multiput(1368.00,156.02)(7.980,-0.507){7}{\rule{4.847pt}{0.122pt}}
\multiput(1368.00,156.27)(57.939,-5.000){2}{\rule{2.424pt}{0.350pt}}
\end{picture}
\vskip 1cm
\begin{center}
\bf{Fig. 1}
\end{center}
\end{document}